\documentclass[a4paper,11pt,oneside]{article}
\usepackage[margin=1in]{geometry}
\usepackage{graphicx}
\usepackage{mathabx}
\usepackage{authblk}
\usepackage{lineno, blindtext}
\usepackage[superscript,biblabel]{cite}
\usepackage{caption}
\usepackage[colorlinks=true,
            linkcolor=blue,
            urlcolor=blue,
            citecolor=blue]{hyperref}
\title{Fermi-liquid transport beyond the upper critical field in superconducting La$_2$PrNi$_2$O$_7$ thin films}

\author[1,2,3]{Yu-Te Hsu\footnote{These authors contributed equally to this work.}\footnote{ythsu@phys.nthu.edu.tw}}
\author[4,5]{Yidi Liu$^*$}
\author[6]{Yoshimitsu Kohama}
\author[7]{Tommy Kotte}
\author[6]{Vikash Sharma}
\author[4,8]{Yaoju Tarn}
\author[4,8]{Bai Yang Wang}
\author[4,8,9,10]{Zhi-Xun Shen}
\author[4,8]{Yijun Yu\footnote{yuyijun@stanford.edu}}
\author[4,8]{Harold Y. Hwang}

\affil[1]{Department of Physics, National Tsing Hua University, Hsinchu 300044, Taiwan}
\affil[2]{Center for Quantum Science and Technology, National Tsing Hua University, Hsinchu 300044, Taiwan}
\affil[3]{Department of Materials Science and Engineering, National Tsing Hua University, Hsinchu 300044, Taiwan}
\affil[4]{Stanford Institute for Materials and Energy Sciences, SLAC National Accelerator Laboratory, Menlo Park, Stanford, CA 94025, United States}
\affil[5]{Department of Physics, Stanford University, Stanford, CA 94305, United States}
\affil[6]{Institute for Solid State Physics, The University of Tokyo, Kashiwa, Chiba 277-8581, Japan}
\affil[7]{Dresden High Magnetic Field Laboratory (HLD-EMFL), Helmholtz-Zentrum Dresden-Rossendorf, Dresden 01328, Germany}
\affil[8]{Department of Applied Physics, Stanford University, Stanford, CA 94305, United States}
\affil[9]{Geballe Laboratory for Advanced Materials, Department of Physics and Applied Physics, Stanford University, Stanford, CA 94305, United States}
\affil[10]{Stanford Synchrotron Radiation Lightsource, SLAC National Accelerator Laboratory, Menlo Park, CA 94025, United States}

\date{}

\begin{document}

\maketitle

\textbf{Unconventional superconductivity typically emerges out of a strongly correlated normal state, manifesting as a Fermi liquid with highly enhanced effective mass or a strange metal with $T$-linear resistivity in the zero-temperature limit. In Ruddlesden-Popper bilayer nickelates $R_3$Ni$_2$O$_7$, superconductivity with a critical temperature ($T_{\rm c}$) exceeding 80 and 40~K has been respectively realised in bulk crystals under high pressure and thin films under compressive strain.
These advancements create new materials platforms to study the nature of high-$T_{\rm c}$ superconductivity, calling for the characterisation of fundamental normal-state and superconducting parameters therein. 
Here we report detailed magnetotransport experiments on superconducting La$_2$PrNi$_2$O$_7$ (LPNO) thin films under pulsed magnetic fields up to 64~T and access the normal-state behaviour over a wide temperature range between 1.5 and 300~K. We find that the normal state of \textcolor{black}{thin-film} LPNO exhibits the hallmarks of Fermi-liquid transport, including $T^2$ temperature dependence of resistivity and Hall angle, and $H^2$ magnetoresistance obeying Kohler scaling. 
Using the empirical Kadowaki-Woods ratio relating the transport coefficient and electronic specific heat, we estimate a quasiparticle effective mass $m^*/m_e \simeq 10$ in \textcolor{black}{thin-film} LPNO, thereby revealing the highly renormalized Fermi liquid state which hosts the high-temperature nickelate superconductivity.
Our results demonstrate that thin-film LPNO follows the same $T_{\rm c}/T_{\rm F}$ scaling observed across a wide variety of strongly correlated superconductors and \textcolor{black}{establish key characteristics of the normal ground state from which the superconductivity in epitaxially strained bilayer nickelates emerges.}}

\section*{Introduction}
The recent discovery of superconductivity at 80~K in (La,Pr)$_3$Ni$_2$O$_{7}$ under high pressure makes the bilayer Ruddlesden-Popper (RP) nickelates the latest material system exhibiting high-temperature superconductivity \cite{sun2023,wang2024}. Despite intense interests in the nature of superconductivity in RP bilayer nickelates \cite{luo2023,hou2023,dong2024,wang2024c,li2025a,liu2023,oh2023,yang2023,lu2024,wang2024b}, key parameters characterising the normal and superconducting state are still missing. This is in large part due to the high-pressure condition required to induce superconductivity, which limits the type of experiments that can be performed to primarily transport under static magnetic field and x-ray diffraction. 
The subsequent realisation of superconductivity in epitaxially strained (La,Pr)$_3$Ni$_2$O$_{7}$ thin films with a critical temperature $T_{\rm c}\approx$~40~K (refs. \cite{ko2025,zhou2025,liu2025}) creates a new material platform to which a broader selection of experimental probes can be employed at ambient pressure, including photoemission \cite{li2025,shen2025}, x-ray spectroscopy \cite{wang2025}, and electron energy-loss spectroscopy \cite{bhatt2025}. 
Moreover, the availability of superconducting (La,Pr)$_3$Ni$_2$O$_{7}$ thin films enables a straightforward implementation of pulsed high-field magnetotransport experiment, which can reveal key information on its superconducting and normal state, including upper critical field ($H_{\rm c2}$), \textcolor{black}{scaling laws in magnetotransport}, and functional form of low-temperature normal-state resistivity $\rho_{\rm n}(T)$.

In superconducting (La,Pr)$_3$Ni$_2$O$_{7}$ above $T_{\rm c}$, $\rho_{\rm xx}(T)$ in bulk crystal and thin film shows an intriguing dichotomy: a putative strange-metal $T$-linear resistivity is found in bulk crystals \cite{sun2023,zhang2024,wang2024} whereas an apparent Fermi-liquid $T$-quadratic resistivity is found in thin films\cite{zhou2025,liu2025}. As the identification of the non-superconducting electronic ground state is critical to understanding the mechanism of electron pairing and the formation of superconducting condensate, it is essential to establish the form of $\rho_{\rm n}(T)$ below $T_{\rm c}$ in the bilayer nickelates. It has been recently found that a partial substitution of La by Pr in (La,Pr)$_3$Ni$_2$O$_{7}$ favours the formation of bilayer structure \cite{wang2024,zhou2025,liu2025}, leading to a higher crystallinity as reflected by a residual resistivity ratio as high as 8 (ref.~\cite{liu2025}), therefore we choose La$_{2}$PrNi$_2$O$_{7}$ (LPNO) thin film as the model platform to study the magnetotransport behaviour of bilayer superconducting nickelates.

Prior magnetotransport experiments on LPNO thin films estimate that $\mu_0H_{\rm c2}$ in the $T=0$ limit is well above 50~T \cite{ko2025,zhou2025,liu2025}, necessitating the use of pulsed field conditions to access the low-$T$ normal state. Here, we report detailed magnetotransport experiment on superconducting LPNO thin films under pulsed magnetic fields up to 64~T. We reveal salient characteristics of the normal-state transport in LPNO, including 1) a $T^2$ resistivity in the field-induced low-$T$ normal state, 2) a $T^2$ temperature dependence of the Hall angle, and 3) an observation of Kohler scaling in normal-state magnetoresistance. These findings provide strong evidence that the underlying electronic ground state in compressively strained LPNO thin film is a Fermi liquid. 
Our measurements also reveal that $H_{\rm c2}(T=0)$ in LPNO thin film exhibits an anisotropy with $\gamma \approx 2.5$
\textcolor{black}{, measured by the ratio between $H_{\rm c2}$ with field applied parallel ($H_{\rm c2, \parallel}$) and perpendicular to the $a$-$b$ plane ($H_{\rm c2, \perp}$), respectively.}
By estimating the magnitude of key normal-state and superconducting parameters using established approaches, we find that the ratio between $T_{\rm c}$ and effective Fermi temperature $T_{\rm F}$ in bilayer RP nickelates exhibits a scaling of $T_{\rm c}/T_{\rm F}\sim 0.01$, as found in a wide variety of strongly-correlated superconductors \cite{uemura2004,matsuura2023,hu2024,cao2018}, thereby hinting at a universal underlying principle governing the $T_{\rm c}$ magnitude in unconventional superconductors.

\section*{Results}
\subsection*{Form of normal-state resistivity below $T_{\rm c}$}
\begin{figure}[hbtp!!!]
\includegraphics[width=1\linewidth]{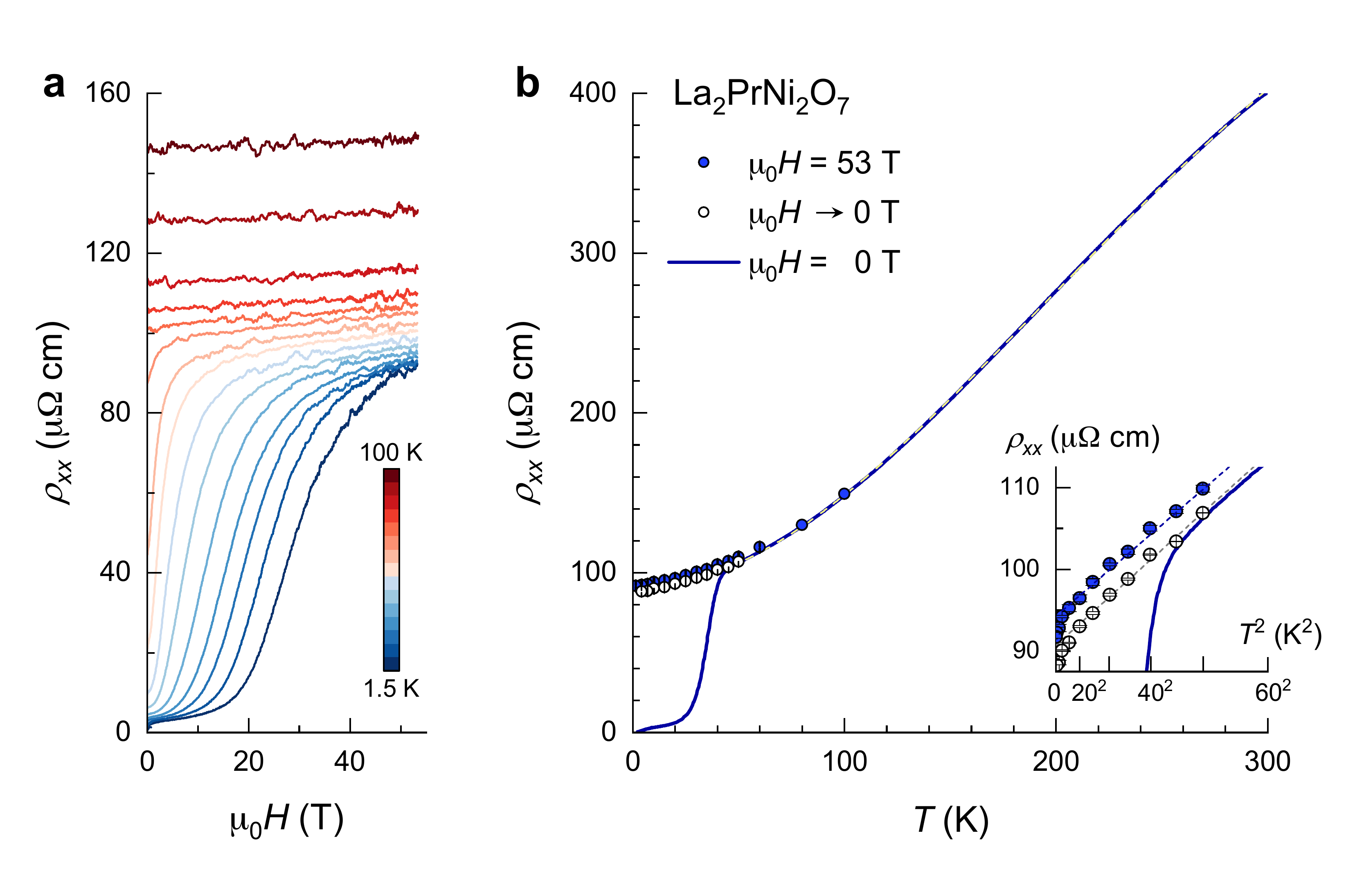}
\caption{\textbf{In-plane resistivity of a superconducting La$_2$PrNi$_2$O$_7$ thin film.}
\textbf{a}, Magnetoresistivity isotherms $\rho_{\rm xx}(H)$ up to 53.4~T measured at the following temperatures: 100, 80, 60, 50, 45, 40, 35, 30, 25, 20, 15, 10, 7.0, 4.1, and 1.5~K.
\textbf{b}, Temperature-dependent resistivity $\rho_{\rm xx}(T)$ measured at \textcolor{black}{0~T (solid line) and 53~T (filled circles), and the extrapolated 0-T values (open circles)}. Grey dash line is a fit to the measured zero-field resistivity above between 50 and 300~K using a parallel resistor model:
$1/\rho(T) = 1/(\rho_0 + A_2 T^2) + 1/\rho_{\rm max}$, 
which finds $\rho_0 = 102~{\rm \mu\Omega~cm}$, $\rho_{\rm max} = 780~{\rm \mu\Omega~cm}$, and $A_2=8.1~\rm{n\Omega~cm~ K^{-2}}$, respectively.
Inset: $\rho_{\rm xx}$ versus $T^2$ below 60~K, showing a $\Delta\rho(T)~\propto~ T^2$ behaviour in the measured 53-T resistivity and extrapolated zero-field resistivity.
\textcolor{black}{For a discussion of the possible origin and impact of two-step transition below $T_{\rm c}$, refer to Supplementary Materials Sec. C.}
}
\label{fig1}
\end{figure}

Figure \ref{fig1}a shows the in-plane resistivity $\rho_{\rm xx}(H,T)$ of a LPNO thin film under pulsed magnetic field up to 53~T. The field is applied parallel to the crystalline $c$-axis i.e. perpendicular to the film surface. A clear transition from a superconducting to a resistive state can be seen below $T_{\rm c}=41$~K \textcolor{black}{(defined as the temperature at which the measured $\rho_{\rm xx}(T)$ falls below $0.9\rho_{\rm n}(T)$ with $\rho_{\rm n}(T)$ obtained by the parallel-resistor fit to the normal-state resistivity; Fig.~\ref{fig1}b}). The temperature dependence of resistivity measured in the field-induced normal state at 53~T and extrapolated to 0~T are plotted in Fig.~\ref{fig1}b. A prominent feature of $\rho_{\rm xx}(T)$ is the positive curvature as $T\rightarrow 0$ and negative curvature as $T\rightarrow 300$~K. As demonstrated in previous work \cite{liu2025}, $\rho_{\rm xx}(T)$ above $T_{\rm c}$ can be well-described using a parallel resistor formula (PRF)\cite{cooper2009}:
\begin{equation}
1/\rho(T) = 1/({\rho_0 + A_2T^2}) + 1/\rho_{\rm max},
\end{equation}
where $\rho_0$, $\rho_{\rm max}$, and $A_2$ are fitting constants 
\textcolor{black}{(For discussion of the physical significance of $\rho_{\rm max}$, refer to Supplementary Materials Sec. E)}.
By suppressing superconductivity with large magnetic field, we find that the field-induced normal-state resistivity below $T_{\rm c}$ closely tracks the PRF fit and shows a clear saturation towards an apparent $T^2$-behaviour as $T \rightarrow 0$ (Fig.~\ref{fig1}b inset). The same behaviour is reproduced in a second sample (Supplementary Materials Sec. I). 
Furthermore, the extrapolated zero-field resistivity exhibits essentially the same $\rho(T)$ form as $\rho_{\rm xx}(\textrm{53~T})$ (see details for zero-field resistivity extrapolation in Methods \textcolor{black}{and Supplementary Materials Sec. D}), demonstrating that the normal-state $\rho_{\rm xx}(T)$ of LPNO exhibits a $T^2$ functional form in the $T=0$ limit, characteristic of a Fermi-liquid ground state.

\subsection*{Hall effect and magnetoresistance}
\begin{figure}[hbtp!!!]
\includegraphics[width=1\linewidth]{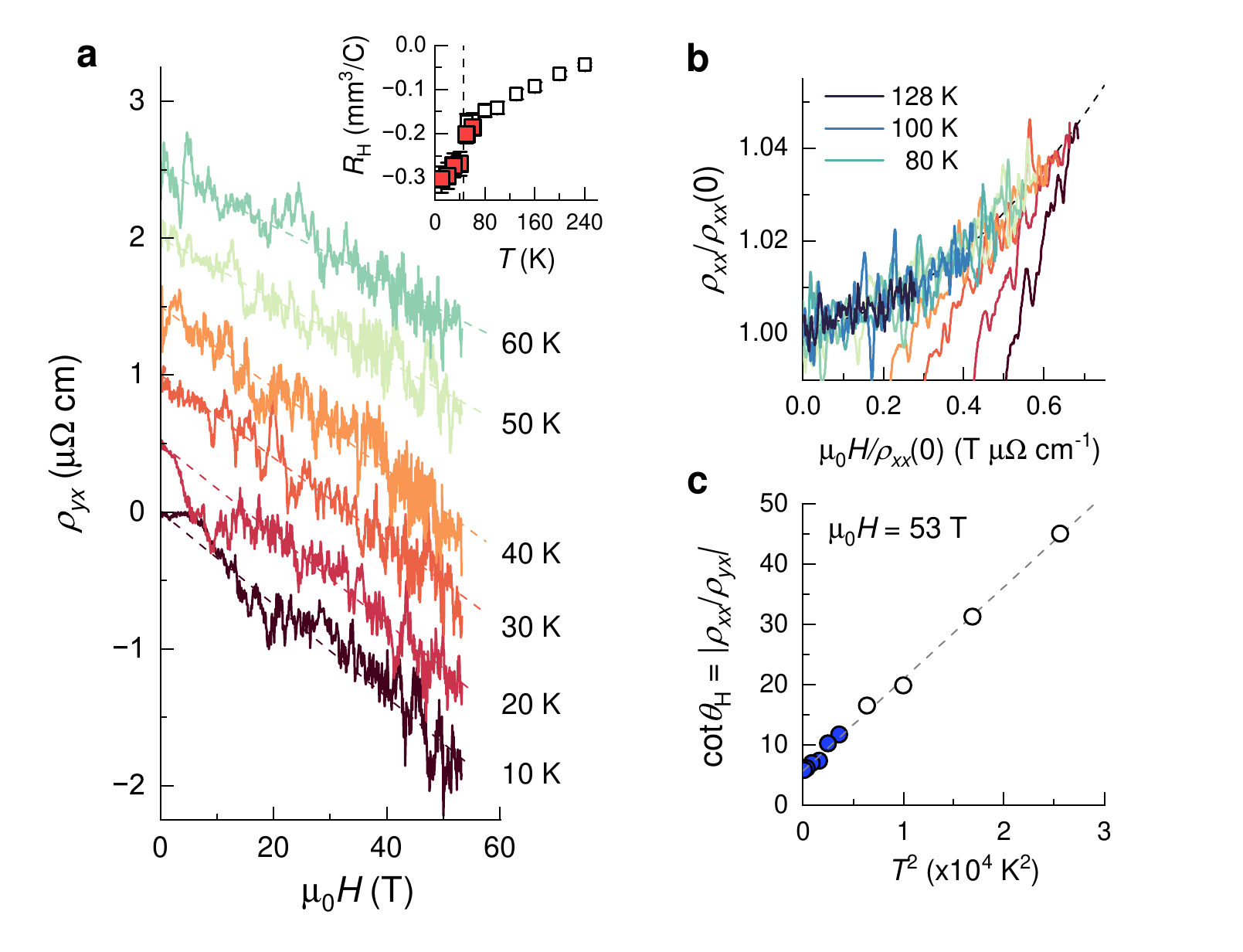}
\caption{\textbf{Hall resistivity, Kohler scaling, and Hall angle in La$_2$PrNi$_2$O$_7$ thin film.}
\textbf{a}, Hall resistivity isotherms $\rho_{\rm yx}(H)$ measured at indicated temperatures. \textcolor{black}{Traces at different temperatures are shifted successively by 1 $\mu\Omega$~cm for clarity.}
Dash lines are fits made to the normal-state $\rho_{\rm yx}$ at high fields using $R_{\rm H}=\rho_{\rm yx} /(\mu_0 H)$.
Inset: Hall coefficient versus temperature $R_{\rm H}(T)$. $R_{\rm H}$ approaches $-0.3$~mm$^3$/C as $T \rightarrow 0$. Vertical dashed line marks $T_{\rm c}$.
\textbf{b}, Normalised magnetoresistance versus magnetic field scaled by zero-field resistivity i.e. $\rho_{\rm xx}/\rho_{\rm xx}(0)$ versus $\mu_0 H/\rho_{\rm xx}(0)$, known as the Kohler plot. Normal-state data measured at indicated temperatures collapses into a single curve following $\rho_{\rm xx}/\rho_{\rm xx}(0)~\propto~(\mu_0 H/\rho_{\rm xx}(0))^2$ shown by the dashed line.
\textcolor{black}{Note that traces of $T$ = 10, 20, 30, and 40~K deviate from the Kohler scaling function at low $\mu_0 H/\rho_{\rm xx}(0)$ until normal-state behaviour is recovered at sufficiently high fields.
$\rho_{\rm xx}$ ($\rho_{\rm yx}$) traces measured below $T_{\rm c}$ are (anti-)symmetrized, whereas the 50- and 60-K data are measured using the positive-polarity trace only, which nonetheless show good agreement with measurement performed at 14~T.}
\textbf{c}, Inverse Hall angle \textcolor{black}{$\cot\theta_{\rm H}=|\rho_{\rm xx}/\rho_{\rm yx}|$} versus $T^2$.
Filled points correspond to measurements at 53~T and open points correspond to extrapolated 53~T values using Hall resistivity measured at 8~T (i.e. $\rho_{\rm yx}(\rm 53~T)$ = $R_{\rm H}(8~\rm{T})\times(53/8)$). Dashed line is a fit using $\cot\theta_{\rm H}=C+BT^2.$
}
\label{fig2}
\end{figure}

Figure \ref{fig2}a shows the Hall resistivity isotherms $\rho_{\rm yx}(H,T)$ measured between 10 and 60~K on the same sample as shown in Fig.~\ref{fig1}. $H$-linear behaviour in $\rho_{\rm yx}(H)$ (within the noise level) is found in the (field-induced) normal state, consistent with the low-field measurements above $T_{\rm c}$ in La$_3$Ni$_2$O$_7$ and LPNO thin films \cite{ko2025, liu2025}. \textcolor{black}{The magnitude of Hall coefficient $R_{\rm H}(T)$ between 10 and 240~K, plotted in Fig.~\ref{fig2}a inset, shows a monotonic increase as $T$ decreases and $R_{\rm H}$ approaches $-0.3$~mm$^3$/C as $T \rightarrow 10$~K.} 
\textcolor{black}{Since the Fermi surface of thin-film LPNO consists of a small electron-like $\alpha$-pocket and a large hole-like $\beta$-pocket \cite{yang2024,li2025,bwang2025}, the negative sign of $R_{\rm H}$ likely reflects the multiband nature of its electrical transport\cite{wang2025a}, thereby precludes the inference of normal-state carrier density using the measured $R_{\rm H}$, as similarly found in the infinite-layer nickelates \cite{lee2023}}.

The form of normal-state magnetoresistance (MR~=~$[\rho_{\rm xx}(H)-\rho_{\rm xx}(0)]/\rho_{\rm xx}(0)$) is examined in Fig.~\ref{fig2}b. The MR magnitude is small ($\lesssim 4\%$ up to 53~T) and follows a $H^2$-behaviour i.e. $\Delta\rho(H)/\rho(0)~\propto~H^2$. We find that $\rho_{\rm xx}(H)/\rho_{\rm xx}(0)$ collapse into a single curve when plotted against $\mu_0H/\rho_{\rm xx}(0)$, known as the Kohler scaling, implying that the MR magnitude is dictated by the carrier mean free path $\ell \sim \omega_{\rm c}\tau$ (i.e. the product of cyclotron frequency $\omega_{\rm c}$ and carrier lifetime $\tau$), as expected for a Fermi-liquid state characterised by a single quasiparticle lifetime \cite{chan2014}. The fact that the normal-state MR~$\propto~H^2$ and obeys Kohler scaling allows us to extract the zero-field normal-state resistivity below $T_{\rm c}$ as shown in Fig.~\ref{fig1}b 
\textcolor{black}{(see Methods and Supplementary Sec. D for extraction details).}
The Fermi-liquid nature of magnetotransport in thin-film LPNO is further supported by the observation of a negligible MR when the magnetic field is applied parallel to the sample surface i.e. $\mathbf{H} \parallel ab$ (Supplementary Materials Fig.~S5) and a $T^2$ dependence of the Hall angle: $\cot\theta_{\rm H} = |\rho_{\rm xx}/\rho_{\rm yx}|~\propto~ (\omega_{\rm c}\tau)^{-1}$ (Fig.~\ref{fig2}c). These features are in stark contrast to the high-$T_{\rm c}$ cuprates within the strange-metal regime, in which a sizeable MR is found with $\mathbf{H}\parallel ab$ (ref. \cite{ayres2021}) and the longitudinal resistivity and Hall angle exhibit distinct $T$-dependence \cite{chien1991}, suggesting that the longitudinal and Hall transport are governed by two distinct carrier lifetimes.
Overall, we find that the normal-state transport characteristics of superconducting LPNO thin film, including $\rho_{\rm xx}(T)$, MR, and  Hall angle, can all be well-described using the standard transport theory for a conventional Fermi-liquid, the key finding of this work.

\subsection*{Upper critical field anisotropy}
\begin{figure}[hbtp!!!]
\includegraphics[width=1\linewidth]{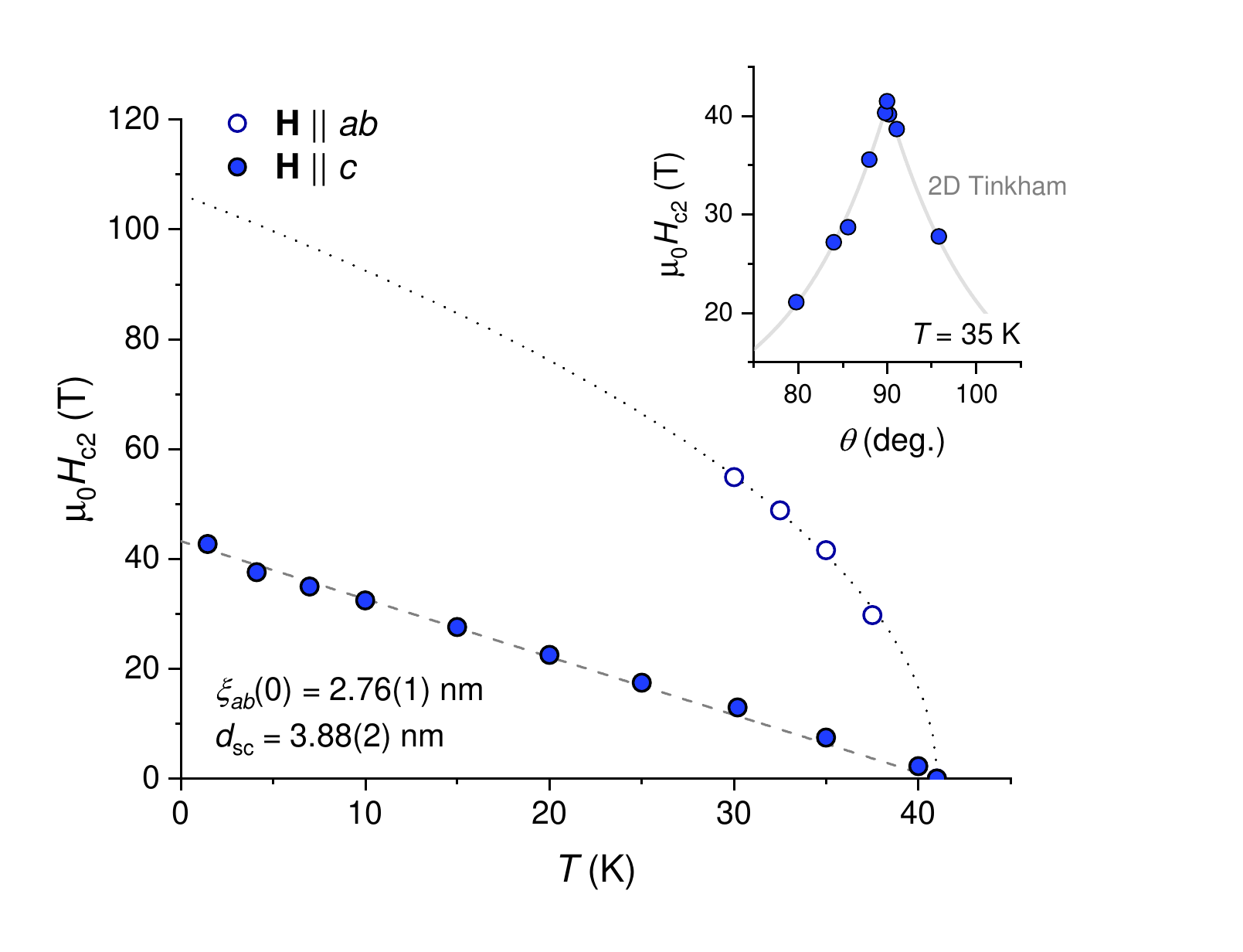}
\caption{\textbf{Upper critical fields of La$_2$PrNi$_2$O$_7$ thin film.}
$\mu_0 H_{\rm c2}$ are extracted using the criterion of $\rho_{\rm xx}(\mu_0H_{\rm c2})/\rho_{\rm xx}$(53~T) = 0.9. Filled points correspond to the configuration that the magnetic field is applied along the film surface normal (i.e. $\mathbf{H}\parallel c$) and open points the configuration that field applied parallel to the film surface (i.e. $\mathbf{H}\parallel ab$). Dash and dotted lines are fits made using the linearised Ginzburg-Landau formulae (see main text).
Inset shows $\mu_0H_{c2}$ measured at 35~K with the field oriented close to the in-plane configuration (i.e. $\theta \simeq 90^{\circ}$). 
Solid line is a fit made using the 2D Tinkham model\cite{tinkham1963}: $\left(\frac{H_{\rm c2}\sin\theta}{H_{\rm c2, \parallel}}\right)^2 +\left|\frac{H_{\rm c2}\cos\theta}{H_{\rm c2, \perp}}\right|=1$.}
\label{fig3}
\end{figure}

Next we extract the upper critical field $H_{\rm c2}$ in both $\mathbf{H}\parallel c$ and $\mathbf{H}\parallel ab$ configuration (i.e. $H_{\rm c2,\perp}$ and $H_{\rm c2,\parallel}$, respectively). Due to the considerable transition width in magnetic field, the choice of different extraction criteria leads to different forms of $H_{\rm c2}(T)$ (Supplementary Materials Fig.~S7). We adopt the 90\% $\rho_{\rm n}$ criterion, in accordance with previous reports \cite{ko2025, liu2025}, and extract $H_{\rm c2}$ using the condition $\rho_{\rm xx}(\mu_0H_{\rm c2})/\rho_{\rm xx}(\rm{53~T)}=0.9$. The $\mu_0H_{\rm c2}(T)$ data\textcolor{black}{, shown in Fig.~\ref{fig3},} is fitted using the linearised Ginzburg-Landau form \cite{wang2021}:
\begin{equation}
    H_{\rm c2,\perp}(T) = \frac{\phi_0}{2\pi\xi_{ab}^2}\left(1-\frac{T}{T_{\rm c}}\right),
\end{equation}
\begin{equation}
    H_{\rm c2,\parallel}(T) = \frac{\sqrt{12}\phi_0}{2\pi\xi_{ab}(0)d_{\rm sc}}\left(1-\frac{T}{T_{\rm c}}\right)^{\frac{1}{2}},
\end{equation}
where $\phi_0$ is the flux quantum, $\xi_{ab}(0)$ is the in-plane coherence length at zero temperature, and $d_{\rm sc}$ is the effective superconducting thickness. 
An anisotropic $H_{\rm c2}$ is found with $\mu_0H_{\rm c2, \perp}(0)=(42.8\pm0.4)$~T and $\mu_0H_{\rm c2, \parallel}(0)=(106\pm1)$~T, corresponding to an anisotropic factor $\gamma = H_{\rm c2, \parallel}/H_{\rm c2, \perp} = 2.48$. 
\textcolor{black}{We further find that the $\mu_0H_{\rm c2}(\theta\simeq 90^\circ)$ data measured at 35~K can be well fitted using the two-dimensional (2D) Tinkham model \cite{tinkham1963} (Fig.~\ref{fig3} inset), pointing to a 2D nature of superconductivity (see Supplementary Sec.~F for details of $H_{\rm c2}$ extraction and model descriptions).}
\textcolor{black}{This $\gamma$ value is comparable to optimally doped YBa$_2$Cu$_3$O$_{7-\delta}$ ($\gamma\approx 2.2$; ref.~\cite{sekitani2004}) and larger than infinite-layer nickelate La$_{0.8}$Sr$_{0.2}$NiO$_2$ ($\gamma\approx1.7$; ref.~\cite{sun2023a}) and iron-based superconductor Ba(Fe,Co)$_2$As$_2$ ($\gamma\approx1.1$; ref.~\cite{kano2009}), implying that the anisotropy in electronic structure of thin-film LPNO is similar to the high-$T_{\rm c}$ cuprates.}
Furthermore, we find the in-plane coherence length $\xi_{ab}(0)=(2.76\pm0.01)$~nm and a superconducting thickness $d_{\rm sc}= (3.88\pm0.02)$~nm, comparable to the film thickness of 5~nm and suggesting a bulk nature of superconductivity. 

Comparing to the previously reported values for La$_3$Ni$_2$O$_7$ and LPNO thin films \cite{ko2025, liu2025}, our extracted $\mu_0H_{\rm c2,\perp}(0)$ is significantly lower. We note that in previous reports, a constant resistivity value at a temperature slightly above $T_{\rm c}$ (e.g. 50~K) is used to define the 90\% $\rho_{\rm n}$ criterion for $\rho_{\rm xx}(H,T)$ measured at $H \ll H_{\rm c2}$. In LPNO thin films, since the normal-state $\rho_{\rm xx}(T)$ follows a $T^2$ behaviour and the $T_{\rm c}\simeq$~40~K is relatively high, the 50-K resistivity value considerably overestimates the true normal-state resistivity at $T \lesssim T_{\rm c}$, thus leading to an overestimation of $H_{\rm c2}(T \lesssim T_{\rm c})$ and $H_{\rm c2}(0)$.
It is also likely that there exists a sample variations in $H_{\rm c2}$ (see Supplementary Materials Fig.~S9), whose underlying cause requires further investigation.
Our current experiment on a LPNO film with an $H_{\rm c2,\perp}$ directly accessible within 53~T provides an accurate quantification of its $\mu_0H_{\rm c2, \perp}(0)$, and allows a crude estimate of the superconducting gap magnitude $\Delta_0 \simeq 6.3$~meV via $\xi_{ab}(0) = \hbar v_{\rm F}/(\pi\Delta_0)$. We caution here that this $\Delta_0$ value should be interpreted as an order-of-magnitude estimate as $v_{\rm F}$ used here is obtained using estimates of $E_{\rm F}$ and $\bar{m}^*$, both of which are associated a sizeable uncertainty, and the level of disorder is assumed to play only a minor role in the extraction of $T_{\rm c}$ and $H_{\rm c2}$ via resistivity measurements.
\textcolor{black}{Nonetheless, we note that recent tunneling experiments \cite{fan2025} have reported $\Delta_0 = (6-20)$~meV for thin-film LPNO, broadly in agreement with our estimate.}

\subsection*{Estimate of effective mass}
It is known that the prefactor of $T^2$-resistivity and electronic specific heat $\gamma_0$ for a variety of correlated electron systems exhibits an empirical relationship\cite{kadowaki1986,jacko2009}: 
\textcolor{black}{$A_2/\gamma_0^2\simeq 10~\mu\Omega~\rm{cm~mol^2~K^2~J^{-2}}$}, known as the Kadowaki-Woods ratio. The high-pressure or epitaxial strain condition required for superconductivity in the bilayer nickelates currently precludes the extraction of $\gamma_0$ via direct calorimetry measurements. Alternatively, by assuming the empirical Kadowaki-Woods ratio of \textcolor{black}{$10~\mu\Omega~\rm{cm~mol^2~K^2~J^{-2}}$} and using the measured $A_2 = (8.1\pm1.2)~\textrm{n}\Omega~\textrm{cm}~\textrm{K}^{-2}$, we infer a $\gamma_0 = (28\pm4)~\textrm{mJ}~\textrm{mol}^{-1}~\textrm{K}^{-2}$ for the LPNO film studied in this work. We further estimate the quasiparticle effective mass using\cite{mackenzie1996}:
\begin{equation}
    \gamma_0 = \gamma\prime\sum_i m_i^*/m_{\rm e},
\end{equation}
where $\gamma\prime = \frac{\pi N_{\rm A}k_{\rm B}^2m_{\rm e}}{3\hbar^2}a^2 = 1.39$~mJ~mol$^{-1}$~K$^{-2}$ ($a=3.756$~\AA~is the in-plane lattice constant) and $m^*_i$ is the effective mass of the $i^{\rm th}$ distinct Fermi pocket in the first Brillouin zone (BZ). 
Given that there are two Ni-O layers in the unit cell of LPNO, two sheets of the Fermi surface are expected in the BZ (i.e. $i=2$) \textcolor{black}{and verified by photoemission experiment \cite{bwang2025}}.
This yields an average effective mass $\bar{m}^* = (10\pm3)~m_e$ for thin-film LPNO, comparable to that found in overdoped cuprates \cite{loram2001,nakamae2003,wang2007,legros2019}. 
\textcolor{black}{We further estimate an effective Fermi temperature $T_{\rm F} = (2330\pm590)$~K using the carrier density and Fermi energy inferred from photoemission data\cite{bwang2025} on thin-film LPNO (see details of the uncertainty estimations in $\bar{m}^*$ and $T_{\rm F}$ in Supplementary Materials Sec. J. and Sec. G). Table~\ref{table} summarises the key normal-state and superconducting parameters of thin-film LPNO extracted from our transport experiment.}

\begin{table}[hbtp!!!]
\begin{center}
\caption{\textbf{Normal-state and superconducting parameters for La$_2$PrNi$_2$O$_7$ thin film discussed in the main text.} Note that the $\Delta_0$ value presented here represents an order-of-magnitude estimate (see main text for discussion).}
\begin{tabular}{cccccc}
\hline
$A_2$ (n$\Omega~$cm$~$K$^{-2}$) & $\xi_{ab}(0)$ (nm) & $\bar{m}^*/m_e$ & $\Delta_0$ (meV) \\
\hline
$8.1 \pm 1.2$ & $2.76\pm0.01$ & $10\pm3$ & 6.3$^{\dag}$ \\
\hline
\end{tabular}
\label{table}
\end{center}
\end{table}

\section*{Discussion}
\begin{figure}[b!!!]
\includegraphics[width=0.75\linewidth]{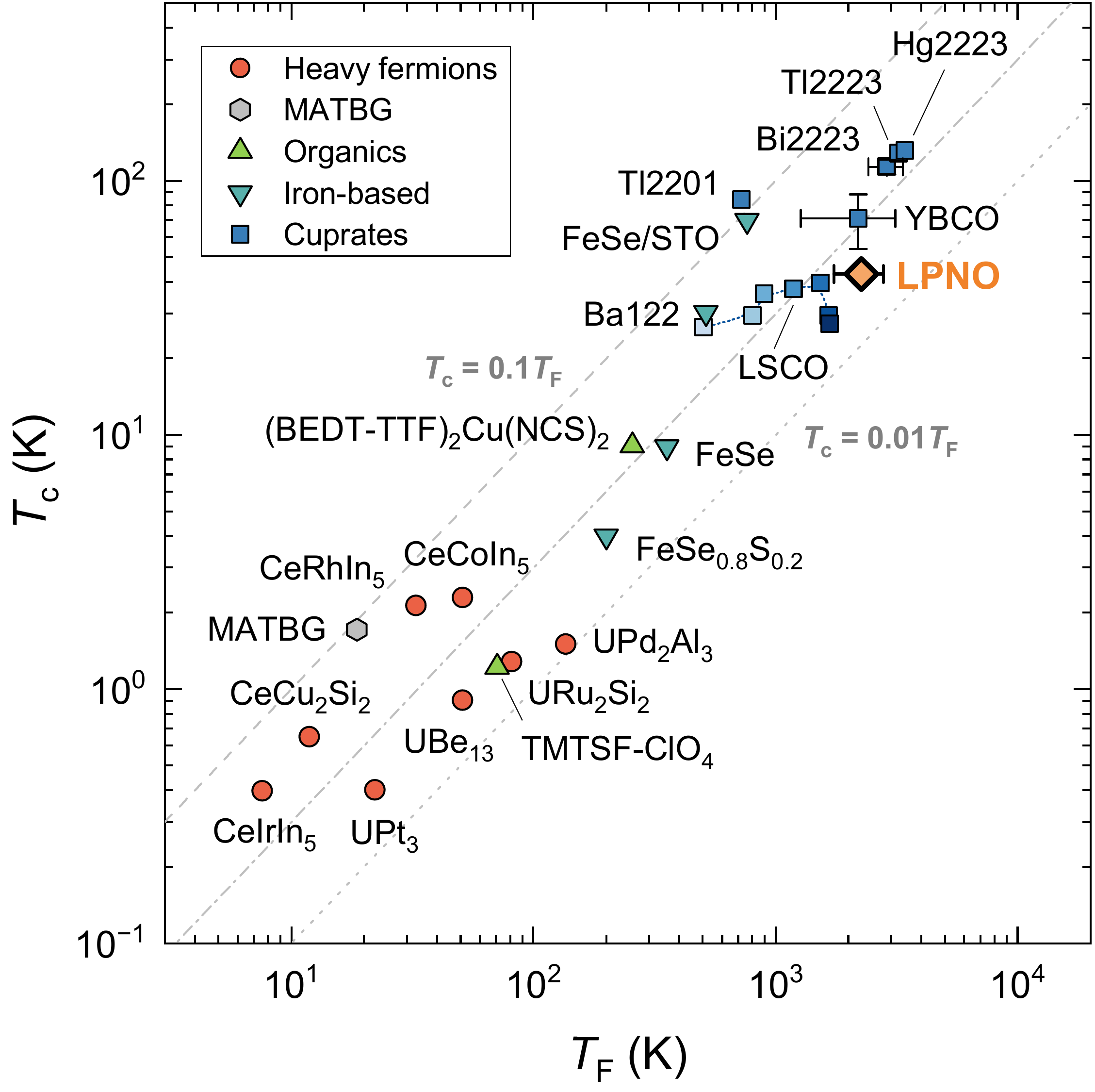}
\centering
\caption{\textbf{Superconducting critical temperature $T_{\rm c}$ versus effective Fermi temperature $T_{\rm F}$ for strongly correlated superconductors.}
The dotted, dotted-dash, and dashed lines indicate $T_{\rm c}/T_{\rm F}$ = 0.01, 0.05, and 0.1, respectively.
For the referenced materials, $T_{\rm F}$ values are extracted assuming a quadratic energy dispersion $E_{\rm F}=\hbar^2k_{\rm F}^2/(2m^*)$ and using experimental data of effective mass and carrier density inferred from specific heat, Hall effect, and quantum oscillation measurements (refs. \cite{uemura2004, cao2018, matsuura2023, hu2024} and references therein) or from penetration depth measurements using $E_{\rm F} = (\hbar^2/2)(3\pi^2)^{2/3}n_{\rm s}^{2/3}/m^*$ (ref.~\cite{uemura1990}).
The series of points in gradient blue illustrates the effect of hole doping on $T_{\rm c}$ and $T_{\rm F}$ of La$_{2-x}$Sr$_x$CuO$_4$ (LSCO) with the darker colour indicating a higher doping level up to $x=0.21$ ($T_{\rm c}=27$~K).
Error bars for the cuprate materials indicate the variations in $T_{\rm c}$ and $T_{\rm F}$ with carrier dopings, and for LPNO the variations in $T^*_{\rm F}$ inferred from the upper and lower limits of $\bar{m}^*$ estimates.
MATBG: magic-angle twisted bilayer graphene;
TMTSF: \textcolor{black}{tetramethyltetraselenafulvalene};
BEDT-TTF: \textcolor{black}{bisethylenedithiol-tetrathiafulvalene};
Ba122: BaFe$_2$(As$_{1-x}$P$_x$)$_2$;
STO: SrTiO$_3$;
YBCO: YBa$_2$Cu$_3$O$_{6+x}$;
Tl2201: Tl$_2$Ba$_2$CuO$_{6+\delta}$;
Tl2223: Tl$_2$Ba$_2$Ca$_2$Cu3O$_{10+\delta}$;
Bi2223: Bi$_2$Sr$_2$Ca$_2$Cu3O$_{10+\delta}$;
Hg2223: Hg$_2$Ba$_2$Ca$_2$Cu3O$_{10+\delta}$;}
\label{fig4}
\end{figure}

It has been noted that, for a wide variety of unconventional superconductors, the ratio between $T_{\rm c}$ and $T_{\rm F}$ share a very similar magnitude i.e. $T_{\rm c}/T_{\rm F} \sim 0.01$ (refs.~\cite{uemura2004, cao2018, matsuura2023, hu2024}). A survey of $T_{\rm c}$ versus $T_{\rm F}$ for notable superconductors, including high-$T_{\rm c}$ cuprates\cite{uemura2004}, iron-based superconductors\cite{matsuura2023}, heavy fermion materials\cite{hu2024}, and carbon-based superconductors\cite{cao2018}, is shown in Fig.~\ref{fig4}. We find that for LPNO the empirical relationship $T_{\rm c}/T_{\rm F} \sim 0.01$ also holds, suggesting a strongly correlated nature of superconductivity in the bilayer RP nickelates and the magnitude of $T_{\rm c}$ is governed by a universal underlying principle independent of material-specific details.

\clearpage

Our high-field experiment demonstrates that the low-$T$ normal state of compressively strained LPNO thin film exhibits the defining characteristics of Fermi-liquid transport. The dichotomy of normal-state resistivity between pressurised single crystal and strained thin film of superconducting bilayer nickelates thus raises a fundamental question of the appropriate electronic ground state description therein. One possible cause for the contrasting normal-state transport is the impact of effective carrier doping\cite{li2025}.
\textcolor{black}{It is well established that in unconventional superconductors with a superconducting dome, the $\rho(T)$ functional form depends strongly on the effective doping level; a predominantly $T$-linear resistivity is often found near the optimal doping level, contrasting a $T^2$ resistivity in the overdoped region on its phase diagram \cite{cooper2009, analytis2014,lee2023,hsu2024,xia2025}.
In such a scenario, our current finding then suggests a possible mapping of epitaxially strained LPNO to the overdoped regime in its putative superconducting dome.
Recent transport experiment on La$_2$PrNi$_2$O$_{7-\delta}$ with a controlled oxygen deficiency $\delta$ (ref.~\cite{wang2025}), however, shows an absence of $T$-linear resistivity with $0.65 \leq \delta \leq0$, implying that an alternative doping strategy is required to explore the putative strange-metal transport in LPNO thin films.}

Another possibility to account for the contrasting $\rho(T)$ is the qualitative different effect between hydrostatic pressure and epitaxial strain, leading to an opposite change in $c$-axis length and possibly critical distinction in the underlying electronic structure.
It has been demonstrated that in epitaxial thin film of La$_{2-x}$Sr$_x$CuO$_4$ grown on LaSrAlO$_4$ near optimal doping ($p=0.15$), $\rho(T)$ evolves from being $T$-linear above $T_{\rm c}$ to being $T$-superlinear \textcolor{black}{(i.e. $\rho(T)\sim~T^n~\textrm{with}~n > 1$)} as the film thickness reduces from 90 to 3~nm \cite{sato2008}, accompanied by a reduction in residual resistivity ratio (RRR=$\rho(\textrm{300~K})/\rho_0$) from 6 to 2. We note that in the case of La$_3$Ni$_2$O$_7$ and LPNO, the bulk crystals showing a $T$-linear resistivity have RRR values of 2.0\textendash2.5 (refs.\cite{sun2023,zhang2024,wang2024}), whereas our thin films show RRR $\approx 4$ (Fig.~\ref{fig1} and Supplementary Materials Fig.~S9). 
It is thus unlikely that the deviation from a $T$-linear resistivity in LPNO thin films is caused by an enhanced level of disorder, and that the distinction in $\rho(T)$ functional form between that found in bulk crystals and thin films reflects a difference in effective carrier doping and/or strain anisotropy 
\textcolor{black}{(see further discussion in Supplementary Materials Sec. H).}

Lastly, we comment on the surprising fact that high-temperature superconductivity emerges out of a Fermi-liquid ground state in LPNO thin films.
While a $T^2$-resistivity is expected for a conventional metallic state as $T\rightarrow 0$, it is generally not expected that such resistivity functional form will hold at high temperatures; instead, $\rho(T)$ in conventional metals typically exhibits a $T$-linear behaviour at intermediate temperature range 
\textcolor{black}{between $T_{\rm D}/3 \lesssim T \lesssim T_{\rm D}$ due to dominant electron-phonon coupling (for bulk La$_3$Ni$_2$O$_7$, a Debye temperature $T_{\rm D} = 383$~K is reported for polycrystals \cite{wu2001}).} In the current case of thin-film LPNO, in no temperature range up to 300~K is a $T$-linear resistivity found, implying that the electron-electron interaction therein is so dominant that the electron-phonon coupling becomes essentially negligible.
This dominant electron-electron coupling is likely to be relevant for understanding the formation of density wave orders in bilayer nickelates \cite{chen2024,chen2024a,khasanov2025}. 
\textcolor{black}{To the best of our knowledge, the only other material with an apparent $T^2$-resistivity up to room temperature is the electron-doped cuprates $R_{2-x}$Ce$_2$CuO$_4$ ($R=$~La, Nd, Pr), whose origin remains to be understood \cite{greene2020}; compared to thin-film LPNO, however, the normal-state resistivity of electron-doped cuprates shows no sign of saturation up to 700~K (ref. \cite{bach2011}), thus the mechanisms behind the high-temperature $T^2$ resistivity in electron-doped cuprates and thin-film LPNO are presumably distinct.}
\textcolor{black}{If the compressively strained LPNO thin films are to be mapped to the overdoped regime of its putative superconducting dome}, it would then be possible to further optimize the $T_{\rm c}$ of RP bilayer nickelates provided a suitable tuning strategy.

\newpage
\section*{Methods}
\subsection*{Thin film growth and device fabrication}
LPNO thin films of approximately 5-nm thickness with a single-unit-cell capping layer of SrTiO$_3$ were grown on SrLaAlO$_4$(001) substrates using pulsed-layer deposition, as detailed in previous work \cite{liu2025}, 
\textcolor{black}{with the crystal structure of thin films confirmed by x-ray diffraction (Supplementary Materials Fig.~S1.}
Gold electrodes of 40-nm thickness with Hall-bar (or van der Pauw) geometry were deposited prior to ozone annealing in a tube furnace \textcolor{black}{(Supplementary Materials Fig.~S2.} Following transport characterisation measurements conducted using $^4$He cryostats, the samples were kept cold during storage at liquid nitrogen temperature and transported to high-field facilities at dry-ice temperature.

\subsection*{Magnetotransport measurements}
Ultrasonically bonded aluminum wires were connected to the measurement probe using 25 $\mu$m-diameter gold wires and room-temperature silver paint (DuPont 4929N). Samples were exposed to ambient conditions for less than two hours between taken out of cryogenic storage and loaded into measurement cryostat. 
Electrical transport measurements \textcolor{black}{using four-probe method} with an AC excitation current of 10~$\mu$A applied along the crystalline $a$-axis were performed in static magnetic fields up to 13~T using a Physical Properties Measurement System by Quantum Design Inc. at a low frequency between $13-30$~Hz, and in pulsed magnetic fields up to 53.4~T and 64.2~T at the International MegaGauss Science Laboratory (IMGSL) in Institute for Solid State Physics, University of Tokyo and Dresden High Magnetic Field Laboratory (HLD-EMFL), Helmholtz-Zentrum Dresden-Rossendorf, respectively. \textcolor{black}{Measurements at the IMGSL are recorded using a National Instruments PXIe-6124 multifunction DAQ device, operating at a sampling rate of 4~MS/s, which applies an AC excitation current of 50~kHz and simultaneously record the resultant AC voltage response of the sample, yielding approximately 2000 data points over the duration of the field pulse.
Measurements at the HLD were conducted using a Stanford Research Systems DS360 function generator (excitation frequency $\sim$3kHz) in combination with a Yokogawa DL850 ScopeCorder. During the field pulse, voltage drops across the sample and a 10~k$\Omega$ series resistor were recorded at a sampling rate of 1~MS/s over 500~ms and subsequently analyzed using a digital lock-in procedure. The excitation current was set to $\approx$10~$\mu$A. 
Magnetic field profiles of the pulse magnets used in this work are shown in Supplementary Material Fig.~S3.}

\subsection*{Extrapolation of zero-field resistivity below $T_{\rm c}$}
The magnetoresistivity data at magnetic fields above which $\rho_{\rm xx}(H)$ shows a positive curvature is fitted to: $\rho(H,T) = \rho(0,T) +\beta H^2$ for extracting the zero-field resistivity $\rho_{\rm xx}(0,T)$ below $T_{\rm c}$. We further adjust the extracted $\rho_{\rm xx}(0,T)$ values to achieve a scaling collapse of the Kohler plot for $\rho_{\rm xx}(H,T)$ measured over the entire temperature range. 
\textcolor{black}{Details of the normal-state resistivity extractions are described in Supplementary Materials Sec. D.}
The adjustment of $\rho_{\rm xx}(0,T)$ required for Kohler scaling is typically less than 1.0~$\mu\Omega~$cm.

\subsection*{Extraction of normal-state parameters}
\textcolor{black}{Electronic specific heat ($\gamma_0$) and average effective mass ($\bar{m}^*$) are calculated using the following equations:
\begin{equation}
\frac{A_2}{\gamma_0^2}=10~\mu\Omega~\textrm{cm}~\textrm{mol}^{2}~\textrm{K}^2~\textrm{J}^{-2},
\end{equation}
\begin{equation}
\gamma_0 = 2\left(\frac{\pi N_{\rm A}k_{\rm B}^2}{3\hbar^2}a^2\right)\bar{m}^*,
\end{equation}
where $N_{\rm A}$ is the Avogadro number, $k_{\rm B}$ the Boltzmann constant, $\hbar$ the reduced Planck constant, and $a$ the in-plane lattice constant \cite{liu2025}.
Effective Fermi temperature ($T_{\rm F}$) is estimated using photoemission data via two approaches as detailed in Supplementary Materials Sec. G. In short, the first approach fits the energy dispersion $E(k)$ of the $\alpha$-band near the Fermi level using a parabolic function and estimate the Fermi energy ($E_{\rm F}$) using the difference between the band bottom and Fermi level. $T_{\rm F}$ and Fermi velocity ($v_{\rm F}$) are then calculated using:
\begin{equation}
T_{\rm F} = E_{\rm F}/k_{\rm B},
\end{equation}
\begin{equation}
v_{\rm F} = \sqrt{2E_{\rm F}/\bar{m}^*}.
\end{equation}
The second approach calculates the average Fermi wavevector ($\bar{k}_{\rm F}$) using a weighted sum of measured $k_{\rm F}$ for the $\alpha$- and $\beta$-pocket by their carrier density:
\begin{equation}
\bar{k}_{\rm F} = \frac{n_\alpha k_{\rm F, \alpha}+n_{\beta}k_{\rm F,\beta}}{n_{\alpha}+n_{\beta}},
\end{equation} 
\begin{equation}
E_{\rm F} = \frac{\hbar^2\bar{k}_{\rm F}^2}{2\bar{m}^*}.
\end{equation}
A good agreement is found between the $T_{\rm F}$ values estimated by these two approaches.
}

\clearpage
\section*{Data availability}
The source data used to create the main figures are presented with this paper. Additional data are available from the corresponding authors upon request.

\section*{Author contributions}
Y.T.H. and Y.Y. conceived the project. 
Y.Y., Y.T.H. and H.Y.H. designed the project.
Y.L., Y.T. and Y.Y. synthesised and characterised the bilayer nickelate thin films under H.Y.H.'s supervision.
Y.T.H., Y.K., T.K. and V.S. performed the high-field magnetotransport measurements.
\textcolor{black}{B.Y.W. and Z.X.S. analysed the photoemission data.}
Y.T.H. analysed the transport data and wrote the manuscript with input from all authors.

\section*{Acknowledgements}
We thank Toni Helm for experimental support with pulsed-field measurements.
We also thank Mark Gibson and Jiarui Li for assistance in sample preparations.
This work is supported by the Yushan Fellow Program (MOE-112-YSFMS-0002-002-P1) and the Center for Quantum Science and Technology (CQST) within the framework of the Higher Education Sprout Project by the Ministry of Education (MOE), Taiwan, and by the National Science and Technology Council, Taiwan (NSTC 113-2112-M-008-044-MY3) (Y.T.H). 
This work is also supported by the US Department of Energy, Office of Science, Basic Energy Sciences, Materials Sciences and Engineering Division under contract no. DE-AC02-76SF00515 (synthesis and transport measurements) as well as SuperC and the Kavli Foundation (in situ transport) (Y.L., Y.T., Y.Y., H.Y.H.), and by Grants-in-Aid for Scientific Research (KAKENHI, grant numbers 22H00104 and 25H00600) from the Japanese Society for the Promotion of Science (JSPS) (Y.K., V.S.).
\textcolor{black}{Part of this work is supported by the US Department of Energy (DOE), Office of Science, Office of Basic Energy Sciences, Materials Sciences and Engineering Division, under contract DE-AC02-76SF00515 (B.Y.W., Z.X.S).}
We further acknowledge support of the HLD at HZDR, member of the European Magnetic Field Laboratory (EMFL) as well as support under the European Union's Horizon 2020 research and innovation programme through the ISABEL project (No. 871106).

\section*{Competing interests}
The authors declare no competing interests.

\end{document}